\documentclass[conference]{IEEEtran}
\IEEEoverridecommandlockouts

\usepackage{amsmath,amssymb,amsfonts,amsthm}
\usepackage{algorithmic}
\usepackage{graphicx}
\usepackage{textcomp}
\usepackage{xcolor}
\usepackage{orcidlink}
\usepackage{xcolor}
\usepackage{siunitx}
\usepackage[nolist]{acronym}
\usepackage{placeins}
\usepackage[caption=false,font=footnotesize]{subfig}
\usepackage{booktabs}
\usepackage{svg}
\usepackage[export]{adjustbox}
\usepackage{multirow}
\usepackage{float}
\usepackage{tabularx,booktabs, tabulary}
\usepackage{lipsum}
\usepackage{adjustbox}
\usepackage[capitalise]{cleveref}
\usepackage{graphicx}
\usepackage{svg}
\usepackage{tikz}
\usepackage{textcomp}
\usepackage{hyperref}
\usepackage{lipsum}

\Crefname{equation}{Eq.}{Eqs.}
\Crefname{figure}{Fig.}{Figs.}
\Crefname{tabular}{Tab.}{Tabs.}

\usepackage[normalem]{ulem}
\usepackage[style=ieee,dashed=false, maxcitenames=1,mincitenames=1,isbn=false,url=true,doi=false,date=year]{biblatex}

\usepackage{amsmath,amssymb,amsfonts}
\usepackage{algorithmic}
\usepackage{graphicx}
\usepackage{textcomp}
\usepackage{xcolor}
\def\BibTeX{{\rm B\kern-.05em{\sc i\kern-.025em b}\kern-.08em
    T\kern-.1667em\lower.7ex\hbox{E}\kern-.125emX}}

\newcommand\copyrighttext{%
  \footnotesize \textcopyright 2025 IEEE. Personal use of this material is permitted.
  Permission from IEEE must be obtained for all other uses, in any current or future
  media, including reprinting/republishing this material for advertising or promotional
  purposes, creating new collective works, for resale or redistribution to servers or
  lists, or reuse of any copyrighted component of this work in other works.}
\newcommand\copyrightnotice{%
\begin{tikzpicture}[remember picture,overlay]
\node[anchor=south,yshift=10pt] at (current page.south) {\fbox{\parbox{\dimexpr\textwidth-\fboxsep-\fboxrule\relax}{\copyrighttext}}};
\end{tikzpicture}%
}

\begin{document}

\title{Automotive Middleware Performance: Comparison of FastDDS, Zenoh and vSomeIP\\
{}
\thanks{This research is accomplished within the project ”AUTOtechagil” (FKZ 01IS22088x).}
}

\author{\IEEEauthorblockN{David Philipp Klüner}
\IEEEauthorblockA{\textit{Chair of Embedded Software} \\
\textit{RWTH Aachen University}\\
Aachen, Germany \\
kluener@embedded.rwth-aachen.de}
\and
\IEEEauthorblockN{Lucas Hegerath}
\IEEEauthorblockA{\textit{Chair of Embedded Software} \\
\textit{RWTH Aachen University}\\
Aachen, Germany \\
hegerath@embedded.rwth-aachen.de}
\and
\IEEEauthorblockN{Amin Dieter Hatib}
\IEEEauthorblockA{\textit{Chair of Embedded Software} \\
\textit{RWTH Aachen University}\\
Aachen, Germany \\
hatib@embedded.rwth-aachen.de}
\and
\IEEEauthorblockN{Stefan Kowalewski}
\IEEEauthorblockA{\textit{Chair of Embedded Software} \\
\textit{RWTH Aachen University}\\
Aachen, Germany \\
kowalewski@embedded.rwth-aachen.de}
\and
\IEEEauthorblockN{Bassam Alrifaee}
\IEEEauthorblockA{\textit{Department of Aerospace Engineering} \\
\textit{University of the Bundeswehr Munich}\\
Munich, Germany \\
bassam.alrifaee@unibw.de}
\and
\IEEEauthorblockN{ Alexandru Kampmann}
\IEEEauthorblockA{\textit{Chair of Embedded Software} \\
\textit{RWTH Aachen University}\\
Aachen, Germany \\
kampmann@embedded.rwth-aachen.de}
}

\maketitle
\copyrightnotice

\begin{acronym}\itemsep0pt
    \acro{E/E}{Electrical/Electronic}
    \acro{IVN}{In-Vehicular Network}
    \acro{CAV}{Connected and Automated Vehicle}
    \acro{CAN}{Controller Area Network}
    \acro{AV}{Automated Vehicle}
    \acro{AD}{Automated Driving}
    \acro{SDN}{Software-Defined Network}
    \acro{VM}{Virtual Machine}
    \acro{SOA}{Service-oriented Architecture}
    \acro{ROS 2}{Robotic Operating System 2}
    \acro{CPS}{Cyber-Physical System}
    \acro{ECU}{Electronic Control Unit}
    \acro{AA}{Adaptive Application}
    \acro{IPC}{Inter-Process Communication}
    \acro{CM}{Communication Management}
    \acro{ROS}{Robot Operating System}
    \acro{TLS}{Transport Security Layer}
    \acro{SM}{State Management}
    \acro{EM}{Execution Management}
    \acro{OEM}{Original Equipment Manufacturer}
    \acro{ARA}{AUTOSAR Runtime for Adaptive Applications}
    \acro{OS}{Operating System}
    \acro{FC}{Function Cluster}
    \acro{FPGA}{Field Programmable Gate Array}
    \acro{GPU}{Graphics Processing Unit}
    \acro{UCM}{Update and Configuration Management}
    \acro{DL}{Deep Learning}
    \acro{AP}{Adaptive Platform}
    \acro{IDS}{Intrusion Detection System}
    \acro{S2S}{Service-to-Signal}
    \acro{MQTT}{Message Queuing Telemetry Transport}
    \acro{OTA}{Over-The-Air}
    \acro{TSN}{Time-sensitive Networking}
    \acro{UDP}{User Datagram Protocol}
    \acro{NUC}{Intel Next-Unit-of-Computing}
    \acro{TCP}{Transmission Control Protocol}
    \acro{OMG}{Object Management Group}
    \acro{QoS}{Quality of Service}
    \acro{DDS}{Data Distribution Service}
    \acro{IP}{Internet Protocol}
    \acro{SDV}{Software-Defined Vehicle}
    \acro{DMA}{Direct Memory Access}
    \acro{HMI}{Human Machine Interface}
    \acro{ML}{Machine Learning}
    \acro{API}{Application Programming Interface}
    \acro{IMU}{Inertial Measurement Unit}
    \acro{UCM}{Update and Configuration Management}
    \acro{HPC}{High-Performance Computer}
    \acro{Wi-Fi}{Wireless Fiber}
    \acro{4G}{4G}
    \acro{HMI}{Human-Machine-Interface}
    \acro{NIC}{Network Interface Card}
\end{acronym}

\begin{abstract}

In this paper, we analyze the performance of modern automotive communication middleware and their interactions with the operating system kernel under various operating conditions.
Specifically, we examine FastDDS, a widely used open-source middleware, the newly developed Zenoh middleware, and vSomeIP, COVESAs open-source implementation of SOME/IP. 
Our objective is to identify the best performing middleware for specific operating conditions and understand their interaction with the kernel to achieve best performance.
To ensure accessibility, we first provide a concise overview of middleware technologies and their fundamental principles. 
We first introduce our performance testing methodology designed to systematically assess middleware performance metrics such as scaling performance, end-to-end latency, and discovery times across multiple message types, network topologies, and configurations.
Then, we present our methodology to examine the kernel latency contribution and the metrics we consider.
Finally, we compare the resulting performance data and present our results in 12 findings.

Our evaluation code and the resulting data will be made publicly available upon acceptance.
\end{abstract}

\begin{IEEEkeywords}
Middleware, \aclp{AV}, vSomeIP, FastDDS, Zenoh, Communication Middlewares
\end{IEEEkeywords}

\section{Introduction}
\label{sec:intro}
The \ac{SDV} paradigm decouples hardware from software and relies on high-performance, dynamically reconfigurable in-vehicle communication \cite{mckinsey_case_nodate, liu_impact_2022, nichitelea_automotive_2019}.
Therefore, the industry is shifting from static, fieldbus-based networks to flexible, Ethernet-centric backbones \cite{vetter_development_2020}.
Automated driving functions further accelerate the transformation: richer sensor suites generate higher data volumes that exceeds the bandwidth provided by traditional in-vehicle networks (IVN) \cite{mckinsey_case_nodate, wu_oops_2021}.

These trends collectively motivate the adoption of novel, software architectures \cite{zhu_requirements-driven_2021, wang_review_2024} to increase flexibility and scalability.

Automotive communication middlewares are at the heart of these software architectures; they enable data exchange across heterogeneous \acp{ECU}, allow for flexible communication topologies, and shield applications from underlying network complexity \cite{autosar_ap_explanation_sw_arch}.

Multiple middlewares, such as FastDDS, vSomeIP and Zenoh, are of potential interest for the automotive domain \cite{kluner_modern_2024, zhang_comparison_2023}.
However, they are based on different communication protocols, communication patterns, and target potentially different application domains \cite{kluner_modern_2024, zhang_comparison_2023}.
How do these differences affect performance, and which one should be preferred for a given software architecture?

To address this question, we evaluate the performance of the three middlewares in this paper. 
We include FastDDS in our evaluation, as it is the default middleware in \ac{ROS 2} and is supported by a large open source community. 
Furthermore, we also include Zenoh in our evaluation, as the newcomer claims higher performance than existing solutions, as well as vSomeIP, an industrially rooted framework, which has been recently made publicly available.
We measure throughput, latency, and reliability across realistic network topologies, using varying parameter settings and message types.
To better understand observed performance characteristics, we further analyze the origin of transmission latency across user-space middleware, kernel networking layers, and the Ethernet transmission layer.

\subsection{Main Contributions}
\label{sec:intro:contrib}
The main contributions of this paper are:
\begin{enumerate} 
    \item An introduction to automotive middlewares and underlying concepts.
    \item Evaluation of their performance in experiments with varying topologies, parameters, and message types.
    \item Evaluation of kernel network stack latency contribution using syscall and kernel function tracing.
    \item Interpretation of results and recommendations for the application of middlewares.
\end{enumerate}

\subsection{Outline}
\label{sec:intro:outline}
The paper is structured as follows:
\Cref{sec:background} provides an introduction to automotive middlewares and distributed systems in vehicles. 
\Cref{sec:method} presents the parameters that we varied in our experiments, our test methodology, and our experimental setup.
In \Cref{sec:results}, we present the results of our experiments and interpret them.
Finally, \Cref{sec:conclusion} presents our conclusion and the applicability of middlewares under varying conditions.

\section{Related Work}
\label{sec:related}
Various authors have studied the performance and latency characteristics of the \ac{ROS 2} framework.
In 2016, \citeauthor{maruyama_exploring_2016} \cite{maruyama_exploring_2016} investigated the performance of \ac{ROS 2} and its predecessor \ac{ROS}. 
Similarly, \citeauthor{kronauer_latency_2021} \cite{kronauer_latency_2021} investigate the latency effects of \ac{ROS 2} in larger, multi-node systems. 
They determined that the main contributing factors to communication latency in \ac{ROS 2} originated with the middleware and message callback. They also showed that latency scales mostly linearly with node count in a given system.

\citeauthor{wu_oops_2021} \cite{wu_oops_2021} examined the performance of \ac{ROS}, \ac{ROS 2} and Cyber for varying parameters, topologies, and message types. They found that local topologies, where the communication does not leave the machine or even the application, perform best.
Furthermore, \cite{zhang_comparison_2023} compares \ac{DDS}, \ac{MQTT} and Zenoh in the context of distributed \ac{ROS 2} systems. 
Their results indicated that the throughput between \ac{DDS} and Zenoh were roughly equal in Ethernet networks, while Zenoh's performance in \ac{Wi-Fi} and \ac{4G} was better. 

In contrast to these papers, we extend the evaluation to automotive use cases by examining additional topologies, message types, and QoS configurations. 
Previous work also does not consider the kernels contribution to latency and therefore offers little inside on its composition.
We are not aware of prior work that evaluated Zenoh and vSomeIP in depth.

\section{Fundamentals}
\label{sec:background}
Before discussing middleware performance, we present the three middlewares we evaluate. We give brief introductions to the core features of FastDDS, vSomeIP and Zenoh as well as their underlying protocols and network stacks.

\subsection{Communication Middlewares}
\label{sec:background:middlewares}
Middlewares are software frameworks that act as an intermediate layer between the \ac{OS} of an \ac{ECU} and automotive applications \cite{maruyama_exploring_2016}. 
They enable communication within \aclp{IVN}, the fieldbus or Ethernet-based networks in vehicles by connecting various applications, often relying on IP-based communication protocols \cite{kluner_modern_2024}. 
Middlewares provide simplified communication patterns, such as Publish-Subscribe and Request-Response, using the network stack of the underlying operating system \cite{henle_architecture_2022}. 

Many middlewares also include discovery features, allowing communication paths to be dynamically established at run-time \cite{eprosima_fastdds_docs}. Some middlewares also support security options for authentication and encryption and \ac{QoS} configurations to ensure reliable message transmission.
We will now provide a brief overview of three popular middleware implementations.

\begin{figure*}[h!]
    \centering
    \includegraphics[width=0.99\textwidth]{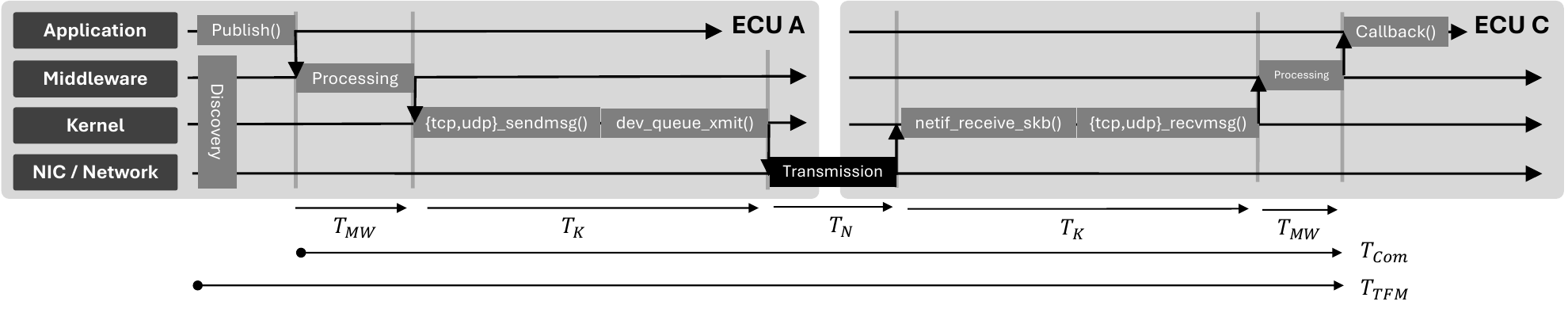}
    \caption{Illustration of our measurement metrics in relation to a transmission process crossing user and kernel space. The metrics are discussed in detail in \cref{sec:method:metrics}.}
    \label{fig:measurements}
\end{figure*}

\subsubsection{SOME/IP}
\label{sec:middleware:someip}

Scalable service-Oriented MiddlewarE over IP (SOME/IP) is an automotive middleware protocol designed specifically for in-vehicle communication. 
The AUTOSAR consortium, composed of \acp{OEM} and tier-one suppliers, standardizes and maintains SOME/IP \cite{autosar_some_ip}.
SOME/IP builds upon conventional transport layer protocols, using \ac{TCP} for reliable transmission and \ac{UDP} for best-effort communication. It supports three interaction models: 
Methods enable remote procedure calls using request-response semantics.
Events implement publish-subscribe communication, allowing services to notify subscribers of state changes. 
Fields represent shared variables exposed through the getter and setter interfaces.

SOME/IP specifies a service discovery mechanism, SOME/IP-SD, as an extension to the base protocol, built on \ac{UDP} multicast communication within the local network segment.
Similarly, SOME/IP-TP \cite{autosar_some_ip_tp} specifies segmentation mechanisms for large messages that do not fit in one single \ac{UDP} packet.
Although SOME/IP itself does not provide built-in encryption or authentication, the AUTOSAR Adaptive Platform specifies security services through the Communication Management (CM) module. 
In this paper, we evaluate vSomeIP, an open-source implementation of the SOME/IP standard by COVESA \cite{vsomeip}. vSomeIP supports the core features of SOME/IP, but does not implement SOME/IP-TP.

\subsubsection{FastDDS}
\label{sec:middleware:fastdds}
FastDDS is an open-source \ac{DDS} implementation by eProsima, based on the DDS standard maintained by the \ac{OMG}.
FastDDS features a large community, as well as integration with \ac{ROS 2}.
\ac{DDS} defines communication patterns and a wire protocol \cite{eprosima_fastdds_docs}.
FastDDS uses a publish-subscribe model where data is associated with topics and accessible to subscribed participants \cite{eprosima_fastdds_docs}.
FastDDS supports discovery and uses multicast \ac{UDP} for participant advertisement.
It also offers dynamic and static discovery options for flexible topology management \cite{eprosima_fastdds_docs}.
FastDDS uses the Real-Time Publish-Subscribe Protocol (RTPS) protocol over \ac{TCP} or \ac{UDP} and supports shared memory communication \cite{eprosima_fastdds_docs}.
The \ac{DDS} standard defines and FastDDS implements extensive \ac{QoS} settings for reliability, message history, and transmission behavior.
Basic configurations are designed for reliability versus best-effort trade-offs \cite{eprosima_fastdds_docs}.
FastDDS also offers extensive security options, such as participant authentication, access control, and traffic encryption defined in the \ac{DDS} security extensions \cite{eprosima_fastdds_docs}.

\subsubsection{Zenoh}
\label{sec:middleware:Zenoh}

Zenoh, released in 2024 as a project under the Eclipse Foundation, is a recently introduced middleware that promises lower latency, reduced transmission overhead, and optimized performance compared to existing solutions. 
Zenoh advertises higher scalability and a design tailored to resource-constrained environments \cite{corsaro_zenoh_2023}
Zenoh structures data as $(\texttt{key},\texttt{value})$  pairs called \textit{resources}.
Each resource is uniquely identified by its key.
Data access is enabled by selectors to match multiple resources (e.g., \textit{/key/*}).
Zenoh provides three primitive operations: \textit{put}, which creates new resources, \textit{delete}, which removes a resource, and \textit{get}, which retrieves the corresponding \texttt{value} \cite{corsaro_zenoh_2023}.
Zenoh defines three different entities. 
\textit{Publishers}: Origin of resources for specific keys or key expressions. 
\textit{Subscribers}: Receivers of resources matching specific keys or expressions. 
\textit{Queryables}: Respond to queries by delivering resources when keys match expressions.

\subsection{Kernel Network Stack}
In the Linux \ac{OS}, the network stack follows a layered architecture \cite{cai_understanding_2021}. 
Messages originating from user-space applications are transferred to the kernel via system calls. 
They are first placed in socket queues and then processed by the transport layer (\texttt{tcp\_send\_message} or \texttt{udp\_send\_message}), which implements the TCP or UDP protocols. These layers rely on the IP layer, using functions such as \texttt{ip\_output} for transmission and \texttt{ip\_deliver} for reception. 
Finally, packets are handed to the \ac{NIC} driver for physical transmission using \texttt{dev\_queue\_xmit()} and \texttt{netif\_receive\_skb()}.

\section{Methodology}
\label{sec:method}

In this section, we describe our methodology to evaluate the performance of the three middlewares. 
We outline our measurement metrics, scenario parameterization, network topologies, and the \acl{QoS} policies configured for the experiments.
In addition, we describe the experimental setup, including the testbed configuration, our testing framework used to execute the experiments, and the process followed to measure, record, and interpret performance data.
 
\subsection{Measurement Metrics}
\label{sec:method:metrics}

We measure middleware performance using the following three metrics, illustrated in \cref{fig:measurements}:
\begin{itemize}
    \item \textbf{Time-To-First-Message ($T_{TFM}$)}: Defined as the time between the synchronized initialization of all communicating applications and the reception of the first message by all recipients reflecting the overall delay in communication startup.
        
    \item \textbf{Communication Time ($T_{Com}$)}: The time between invoking a message transmission,  \texttt{publish}, and the message’s reception by the receiving application in steady state after $T_{TFM}$.
    
    \item \textbf{Communication Time Components ($T_{\{K,M\}}^{\{R,S\}}$)}: The fraction of $T_{Com}$ spent on the middleware ($M$) or the kernel ($K$) on the sender ($S$) or receiver ($R$) side.

    \item \textbf{Ethernet Layer Time ($T_{N}$)}: The portion of $T_{Com}$ spent on physical transmission and NIC. Due to driver variations per device, we cannot distinguish after the handoff to the \ac{NIC} driver.
    
    \item \textbf{Throughput Factor ($S$)}: Defined as $S = \frac{T_{\text{Com}}}{S_{\text{Mes}}}$, this metric relates communication time to message size and reflects throughput scaling with increasing data volumes.

\end{itemize}
\subsection{Experiment Parametrization}
\subsubsection{Message Types}

As outlined by \citeauthor{wu_oops_2021}, common sensors in automated vehicles include \acp{IMU}, cameras, lidars and radars \cite{wu_oops_2021}. To ensure comparability with their findings, we use four sensor types (\textit{IMU, Radar, Lidar, Camera}) as test cases, as detailed in \cref{tab:datasets}.

\begin{table}[bp]
\centering
\caption{Message Types and their respective size and frequency.}
\begin{tabular}{l llll}
\toprule
Metric    & IMU  & Camera & Lidar & Radar \\
\midrule
Frequency {[}Hz{]}      & 200  & 50     & 20    & 20    \\
Message Size {[}kB{]} & 1  & 4000   & 4000  & 10  \\
Throughput {[}MB/s{]} & 0.2  & 200   & 80  & 0.2  \\
\bottomrule
\end{tabular}

\label{tab:datasets}
\end{table}

\subsubsection{Network Topologies}

\begin{figure}[tbp]
    \centering
    \includegraphics[width=0.47\textwidth]{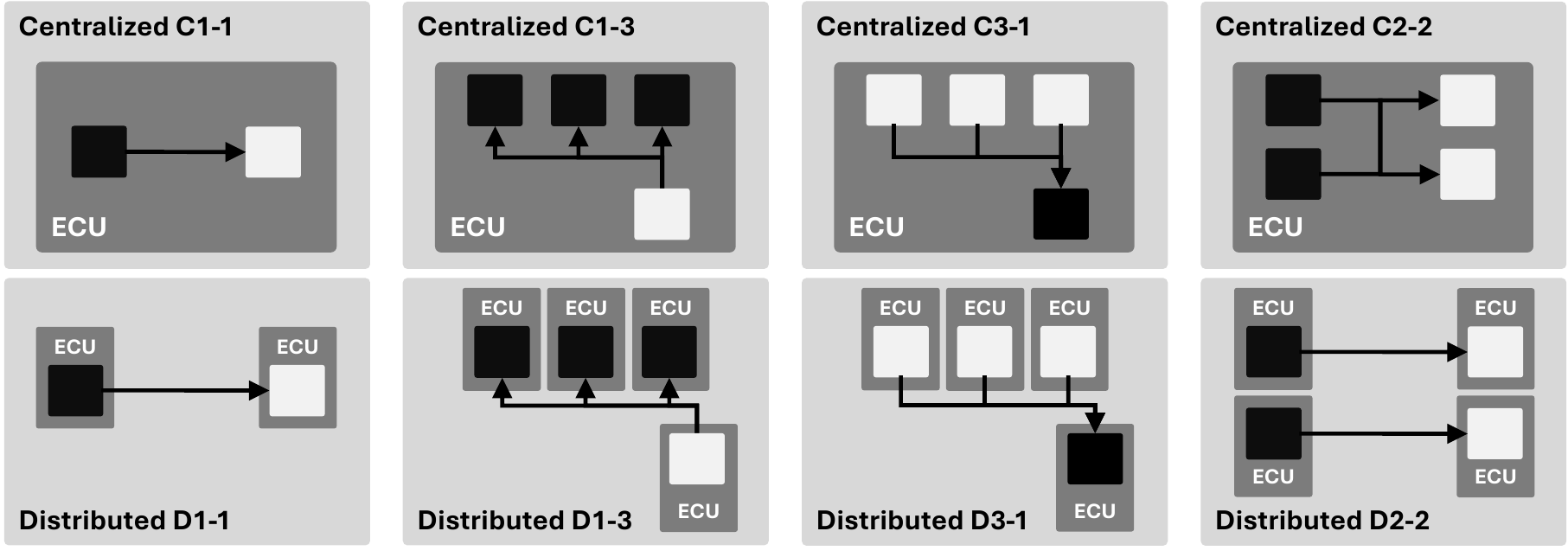}
    \caption{Illustration of all network configurations. Black rectangles denote sending applications, while grey rectangles indicate receiving applications. Arrows represent directed communication between applications.}
    \label{fig:topologies}
\end{figure}

Sensor data processing in automated vehicles can be performed centrally on a vehicle computer or distributed among zone controllers, depending on the vehicle’s \ac{E/E} architecture \cite{zhu_requirements-driven_2021}. 
We evaluate two network topologies: a \textit{centralized} topology, where all applications run on a single platform using efficient \ac{IPC}, and a \textit{distributed} topology, connected by Ethernet-based \ac{IVN}. In both topologies, we consider four combinations of sending and receiving applications: X1-1, X2-2, X3-1, and X1-3, as shown in \cref{fig:topologies}.

\subsubsection{Quality of Service}

Many middleware solutions support configurable \ac{QoS} settings, enabling applications to adjust their behavior \cite{eprosima_fastdds_docs}. Typically, middlewares offer at least two communication modes: best-effort ($BE$) and reliable ($RE$). In best-effort communication, message loss is tolerated in order to reduce overhead, while reliable communication ensures message delivery.
To assess the impact of these configurations, we evaluated one best-effort and one reliable setup for each middleware.
FastDDS provides extensive \ac{QoS} configurability, but to limit the scope of this evaluation, we choose two configurations.
In our evaluation, the reliable setup uses \texttt{keep\_all} durability, \texttt{reliable} reliability, and \texttt{transient\_local} locality, while the best-effort setup uses \texttt{keep\_last} durability, \texttt{best\_effort} reliability, and \texttt{volatile} locality. In contrast, vSomeIP and Zenoh offer limited \ac{QoS} options, relying on the underlying transport protocol to guarantee message delivery. 
Consequently, we configured both Zenoh and vSomeIP to use TCP for reliable communication and UDP for best-effort transmission.

\begin{figure*}[!htb]
    \centering
    
    \subfloat[]{
        \includeinkscape[width=0.98\textwidth]{BOX-comm-times-overview-dist_svg-tex.pdf_tex}
    }
    
    \vspace{0.1ex}
    \subfloat[]{
        \begin{adjustbox}{width=0.7\textwidth,center}
            \begin{tabular}{lll|llll|llll}
                \toprule
                 \textbf{Reliabilty} & \textbf{Middleware} & \textbf{Message Type} & \multicolumn{4}{c|}{\textbf{Max [ms]}} & \multicolumn{4}{c}{\textbf{Median [ms]}} \\
                 &  &  & D1-1 & D1-3 & D2-2 & D3-1 & D1-1 & D1-3 & D2-2 & D3-1 \\
                \midrule
                \multirow[t]{3}{*}{\textbf{Best-Effort}} & FastDDS & IMU & 51.624 & 60.591 & 55.469 & 52.324 & 51.624 & 60.372 & 49.400 & 51.502 \\
                \cline{2-11}
                 & Zenoh & IMU & \textbf{11.783} & \textbf{57.199} & \textbf{14.859} & \textbf{45.706} & \textbf{11.783} & \textbf{12.073} & \textbf{12.645} & \textbf{41.789} \\
                \cline{2-11}
                 & vSome/IP & IMU & \textcolor{gray}{98.443} & \textcolor{gray}{2664.442} & \textcolor{gray}{466.489} & \textcolor{gray}{350.766} & \textcolor{gray}{98.443} & \textcolor{gray}{1051.316} & \textcolor{gray}{294.797} & \textcolor{gray}{350.766} \\
                \bottomrule
            \end{tabular}
            \label{tab:ttfm}
        \end{adjustbox}
    }
    
    \caption{ (a) Scatter plots of $T_{Com}$ samples by topology for the IMU message type and each middleware. 
    (b) Table of $T_{TFM}$ by topology for IMU messages and all middlewares. Maximums for each message type and topology configuration are highlighted in gray, while minimums in each configuration are highlighted in black.}
    \label{fig:disc_table_and_tcom_scatter}
\end{figure*}

\subsection{Experimental Setup}

We conducted the middleware evaluation using a testbed designed by \citeauthor{kampmann_agile_2020} \cite{kampmann_agile_2020}. The testbed consists of four Intel NUCs, each equipped with an Intel Core i5-7260U processor, 8 GB of DDR4 RAM, and an M.2 SSD, interconnected through a central switch using 1 Gbit full-duplex Ethernet. For distributed topologies, we assigned each application to a separate NUC, while in centralized configurations, all applications ran on a single NUC. To synchronize the time between devices, we use the Precision Time Protocol (PTP).  
We automatically performed 142 separate experiments on the presented configurations.

\subsection{Application and Kernel Instrumentation}

We recorded transmission times using LTTng, an open source tracing framework for Linux, with tracepoints instrumented in the middleware and application layers for accurate timing. We also used LTTngs interface for kprobes, to trace the invokation of kernel functions in the network stack. 

In particular, we traced the following kernel symbols:

\begin{enumerate}
    \item \texttt{\{udp,tcp\}\_sendmsg()}/\texttt{\_recvmsg()}: UDP/TCP transport layer entry functions and receive functions.
    \item \texttt{dev\_queue\_xmit()}/\texttt{netif\_receive\_skb()}: Functions we defined as boundary between kernel and network device driver. \texttt{dev\_queue\_xmit()} is invoked shortly before addition of the packet to the NICs TX ring. On the receiving side,  \texttt{netif\_receive\_skb()} is invoked almost directly after the \ac{NIC} has placed a new packet via \ac{DMA} in system memory.
\end{enumerate}

\section{Results}
\label{sec:results}
This section presents the results of our experiments in 12 findings. 
First, we discuss results related to Time-To-First-Message $T_{TFM}$, then communication latency $T_{Com}$. Next, message loss and kernel interactions are discussed. Finally, resource usage and scaling behavior are presented. We omitted two configurations from our test set as they did not yield meaningful results: 
\begin{itemize}
    \item vSomeIP best-effort for configurations larger than our IMU message size: Since vSomeIP does not implement SOME/IP-TP, best-effort transport using UDP is limited to messages that fit within a single UDP payload.
    \item Zenoh reliable: The reliable configuration currently only acts as a decorator, but does not affect message transmission. 
\end{itemize}

\subsection{Time-To-First-Message Findings}
\label{sec:results:discovery}
To understand how quickly a middleware becomes operational after being started, we measure the $T_{TFM}$ as a proxy for discovery time before each $T_{Com}$ experiment.
\cref{fig:disc_table_and_tcom_scatter} a) shows the results of our $T_{TFM}$ experiments for the selected middlewares. 
The following section presents our findings based on this data.

\subsubsection{\textbf{Finding 1: Zenoh consistently has the fastest discovery, closely followed by FastDDS}}
\label{sec:results:finding_1}

In general, in all topologies and configurations, Zenoh achieves a narrowly better or similar performance to FastDDS.
Examining the D1-1 topology as an example, Zenoh completes discovery and message transmission in \SI{11.631}{\milli\second} while FastDDS requires \SI{51.624}{\milli\second} for the same task.
The same trend continues in other topologies, with Zenoh having a lower $T_{TFM}$ than FastDDS.
Zenohs $T_{TFM}$ on average remains similar for most topologies, only in the D3-1 topology does Zenohs $T_{TFM}$ increase to a median of \SI{41.789}{\milli\second}.
FastDDS $T_{TFM}$ remains roughly consistent in the median in the interval of \SIrange{49.400}{60.372}{\milli\second}.

\subsubsection{\textbf{Finding 2: vSomeIP's discovery performs poorly in comparison to FastDDS and Zenoh}}
\label{sec:results:finding_2}

With a median time of \SI{98.443}{\milli\second} to the first message transmission in the D1-1 topology, vSomeIP is 8.3 times slower compared to Zenoh. Its performance in other topologies remains poor, especially in the D1-3 topology, where a median of \SI{1051.316}{\milli\second} is required for first message transmission. 

Particularly notable is the poor scaling to larger topologies, as vSomeIP showed significantly worse performance in topologies other than D1-1, while FastDDS's $T_{TFM}$ remains very similar across all topologies.

\subsection{Communication Latency}
\label{sec:results:latency}
To understand how fast middlewares transmit messages after reaching steady state, we measure $T_{Com}$ in a series of experiments.
\cref{fig:tcom_box_and_tcom_table} a) and \cref{fig:tcom_box_and_tcom_table} b) show the raw results of our $T_{Com}$ experiments for all middlewares. 
The following section presents our findings based on this data.

\subsubsection{\textbf{Finding 3: For communication latency, Zenoh and FastDDS best-effort perform similarly in most circumstances, vSomeIP as a distant third}}
\label{sec:results:finding_3}

\begin{figure*}[!htbp]
    \centering
    
   \subfloat[]{
        \includeinkscape[width=0.99\textwidth]{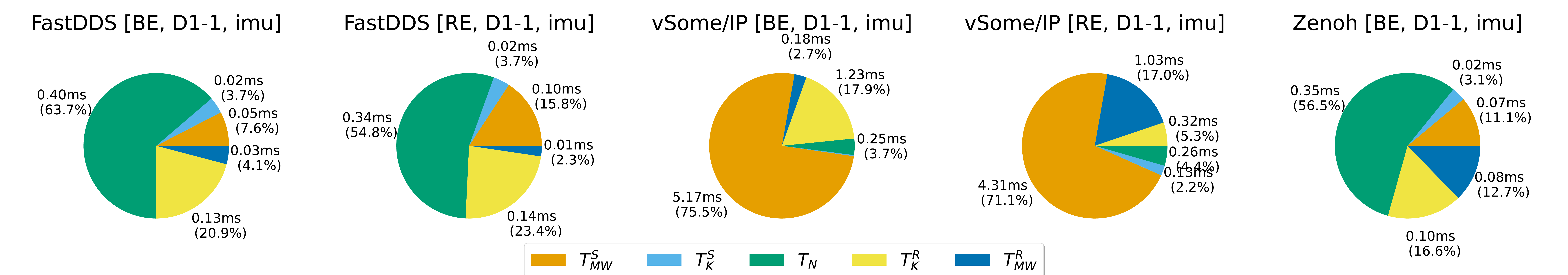}
        \label{fig:kernel_pi}
    }
    \vspace{0.1ex}
    \subfloat[]{
        \begin{adjustbox}{width=0.98\textwidth,center}
        \begin{tabular}{lll|llllllll|llllllll}
            \toprule
             \textbf{Reliabilty} & \textbf{Middleware} & \textbf{Message Type} & \multicolumn{8}{c|}{\textbf{Max [ms]}} & \multicolumn{8}{c}{\textbf{Median [ms]}} \\
             &  &  & C1-1 & C1-3 & C2-2 & C3-1 & D1-1 & D1-3 & D2-2 & D3-1 & C1-1 & C1-3 & C2-2 & C3-1 & D1-1 & D1-3 & D2-2 & D3-1 \\
            \midrule
            \multirow[t]{9}{*}{\textbf{Best-Effort}} & \multirow[t]{4}{*}{FastDDS} & Camera & \textbf{1.995} & \textbf{5.092} & \textbf{5.998} & \textbf{22.137} & 41.531 & \textbf{103.899} & \textbf{122.597} & - & \textbf{1.714} & \textbf{3.482} & \textbf{3.183} & \textbf{2.514} & \textbf{35.367} & 103.244 & 69.685 & - \\
             &  & IMU & \textbf{0.187} & \textbf{0.460} & 0.452 & 0.664 & \textbf{0.572} & \textbf{1.542} & \textbf{1.930} & \textbf{2.307} & \textbf{0.088} & 0.187 & \textbf{0.092} & \textbf{0.069} & \textbf{0.487} & \textbf{1.180} & \textbf{1.019} & \textbf{0.669} \\
             &  & Lidar & \textbf{2.262} & \textbf{4.747} & \textbf{6.222} & \textbf{17.850} & \textbf{35.754} & \textbf{120.733} & \textbf{70.736} & - & \textbf{1.713} & \textbf{3.508} & \textbf{3.327} & \textbf{2.560} & \textbf{35.416} & 103.300 & \textbf{69.749} & - \\
             &  & Radar & \textbf{0.243} & \textbf{0.734} & 0.688 & \textbf{0.499} & \textbf{0.977} & \textbf{2.104} & \textbf{2.184} & \textbf{2.300} & \textbf{0.161} & \textbf{0.356} & \textbf{0.159} & \textbf{0.126} & \textbf{0.722} & \textbf{1.714} & \textbf{1.327} & \textbf{0.893} \\
            \cline{2-19}
             & \multirow[t]{4}{*}{Zenoh} & Camera & 4.167 & 16.173 & 14.715 & 22.922 & 43.836 & 126.702 & \textcolor{gray}{156.686} & \textcolor{gray}{197.011} & 3.774 & 11.077 & 9.022 & 6.050 & 38.528 & \textbf{92.237} & \textcolor{gray}{69.992} & \textbf{65.983} \\
             &  & IMU & 0.276 & 0.527 & 0.645 & \textcolor{gray}{0.991} & 3.246 & 3.879 & 3.815 & 3.733 & \textcolor{gray}{0.132} & \textbf{0.164} & 0.155 & 0.122 & 3.171 & 3.396 & 1.383 & 1.819 \\
             &  & Lidar & 4.543 & 13.818 & 14.759 & 19.567 & \textcolor{gray}{60.628} & 205.260 & 115.520 & \textbf{122.491} & 3.773 & 10.684 & 8.471 & 5.988 & 38.545 & \textbf{92.262} & 70.348 & \textbf{38.970} \\
             &  & Radar & 0.325 & \textcolor{gray}{0.879} & \textcolor{gray}{0.841} & 0.926 & 3.620 & 4.038 & 4.007 & 3.934 & 0.251 & 0.575 & \textcolor{gray}{0.268} & 0.197 & 3.401 & 3.673 & 1.586 & 1.937 \\
            \cline{2-19}
             & vSome/IP & IMU & \textcolor{gray}{0.404} & 0.654 & 0.448 & 0.609 & \textcolor{gray}{8.038} & 8.769 & 8.596 & 8.952 & 0.119 & 0.298 & 0.174 & 0.234 & 7.800 & 7.904 & \textcolor{gray}{7.687} & 8.469 \\
            \midrule
            \multirow[t]{8}{*}{\textbf{Reliable}} & \multirow[t]{4}{*}{FastDDS} & Camera & 24.637 & 8.351 & \textcolor{gray}{82.061} & 28.090 & \textbf{37.874} & 204.942 & - & - & 13.693 & 6.428 & 6.269 & 6.029 & 37.412 & 105.713 & - & - \\
             &  & IMU & 0.228 & \textcolor{gray}{166.074} & \textcolor{gray}{28.845} & \textbf{0.591} & 0.779 & \textcolor{gray}{192.084} & \textcolor{gray}{17.621} & \textcolor{gray}{67.510} & 0.110 & \textcolor{gray}{22.492} & \textcolor{gray}{22.483} & 0.104 & 0.645 & \textcolor{gray}{177.620} & 1.807 & \textcolor{gray}{60.554} \\
             &  & Lidar & 24.667 & 7.572 & \textcolor{gray}{46.844} & \textcolor{gray}{169.033} & 38.264 & \textcolor{gray}{205.899} & - & - & 13.711 & 6.424 & 6.275 & \textcolor{gray}{134.943} & 37.500 & 105.741 & - & - \\
             &  & Radar & 0.333 & 0.735 & 0.795 & \textcolor{gray}{102.754} & 1.221 & 2.457 & 2.283 & 2.354 & 0.196 & 0.396 & 0.197 & \textcolor{gray}{52.118} & 0.933 & 1.864 & 1.379 & 1.047 \\
            \cline{2-19}
             & \multirow[t]{4}{*}{vSome/IP} & Camera & \textcolor{gray}{30.886} & \textcolor{gray}{50.161} & 36.698 & \textcolor{gray}{53.236} & \textcolor{gray}{60.357} & \textcolor{gray}{420.438} & 130.088 & \textbf{134.185} & \textcolor{gray}{22.979} & \textcolor{gray}{29.000} & \textcolor{gray}{24.279} & \textcolor{gray}{34.488} & \textcolor{gray}{56.225} & \textcolor{gray}{123.425} & \textbf{53.865} & \textcolor{gray}{74.281} \\
             &  & IMU & 0.242 & 0.707 & \textbf{0.316} & 0.724 & 8.038 & 8.591 & 11.298 & 8.610 & 0.127 & 0.298 & 0.136 & \textcolor{gray}{0.241} & \textcolor{gray}{7.817} & 7.936 & 5.612 & 6.047 \\
             &  & Lidar & \textcolor{gray}{25.380} & \textcolor{gray}{51.454} & 41.857 & - & 60.016 & 130.974 & \textcolor{gray}{186.114} & \textcolor{gray}{170.701} & \textcolor{gray}{23.685} & \textcolor{gray}{39.580} & \textcolor{gray}{27.260} & - & \textcolor{gray}{56.381} & \textcolor{gray}{123.154} & \textcolor{gray}{89.202} & \textcolor{gray}{69.532} \\
             &  & Radar & \textcolor{gray}{0.356} & 0.844 & \textbf{0.389} & 0.819 & \textcolor{gray}{8.358} & \textcolor{gray}{8.879} & \textcolor{gray}{57.235} & \textcolor{gray}{57.418} & \textcolor{gray}{0.255} & \textcolor{gray}{0.608} & 0.221 & 0.329 & \textcolor{gray}{8.098} & \textcolor{gray}{8.218} & \textcolor{gray}{8.408} & \textcolor{gray}{7.183} \\
            \bottomrule
            \end{tabular}
            \end{adjustbox}
    }
    \caption{(a) Pie charts of kernel impact metrics $T_{\{K,MW\}}^{\{R,S\}}$ for 500 messages in a D1-1 topology. (b) Table of Median and Max $T_{Com}$ values by topology for each reliability, middleware and message type configuration. Maximums for each message type and topology configuration are highlighted in \textcolor{gray}{gray}, minimums in \textbf{black}.}
    \label{fig:tcom_box_and_tcom_table}
\end{figure*}

 In our D1-1 experiments, FastDDS generally outperforms Zenoh and vSomeIP with a mean of \SI{0.487}{\milli\second} for IMU message transmission, compared to \SI{3.170}{\milli\second} and \SI{7.817}{\milli\second}, respectively.  
The differences in D1-1 communication between Zenoh and FastDDS amount to a median of \SI{2.683}{\milli\second} per transmission, while vSomeIP’s performance is particularly weak compared to the other two approaches.
These results also holds for maximum $T_{Com}$ values.
vSomeIP doubles the $T_{Com}$ in contrast to Zenoh for small messages at \SI{7.817}{\milli\second}, and performs similarly poorly for larger Camera messages, as shown in \cref{fig:tcom_box_and_tcom_table}~b).
However, in other topologies and at higher data rates, the picture is more complex, as Zenoh and vSomeIP perform better than FastDDS. 

\subsubsection{\textbf{Finding 4: Zenoh performs well in D1-3, D3-1 and D2-2 topologies at high data rates}} 
\label{sec:results:finding_4}
In our D1-3, D3-1, and D2-2 experiments and large data rates, Zenoh performs at least as well or better than FastDDS in both best-effort and reliable configurations.
However, these results do not hold for maximum $T_{Com}$ values.
As expected, larger message sizes lead to increased latency for all middlewares.
Zenohs median $T_{Com}$ is least influenced by the increase in data rates in these topologies. 
This effect is prominently shown in the D1-3 Camera experiment, where FastDDS required a median of \SI{103.244}{\milli\second} for message transmission, while Zenoh managed to transmit the message with a median $T_{Com}$ of \SI{92.237}{\milli\second}. 
With a transmission delay of \SI{123.425}{\milli\second} vSomeIP reliable was significantly slower than the other solutions. 
As vSomeIP does not implement the Transport Protocol (TP) extension which would allow UDP frame segmentation, we can only use the vSomeIPs reliable configuration using TCP transport in this case.

\subsubsection{\textbf{Finding 5: Evaluation of the impact of QoS could not be meaningfully conducted}}
\label{sec:results:finding_5}
An evaluation of the impact of QoS settings was not comprehensively possible, as only FastDDS could be tested in reliable and best-effort configurations for all message types.
To our knowledge, Zenoh's reliable configuration currently functions only as a decorator without affecting transmission behavior, therefore evaluation would not have been meaningful.
Similarly, vSomeIPs best-effort mode using UDP could only be tested for messages smaller than a single UDP frame, while reliable TCP communication is required for larger messages. 
FastDDS, although supporting both reliable and best-effort modes, produced overall inconclusive results. 
Specifically, FastDDS exhibited a notable, reproducible latency spike of \SI{177.620}{\milli\second} in the median for reliable IMU message transmission, in contrast to a median best-effort $T_{Com}$ of \SI{1.180}{\milli\second}.

\subsubsection{\textbf{Finding 6: FastDDS and Zenoh benefit strongly from \ac{IPC} communication}} 
\label{sec:results:finding_6}
The results shown in \cref{fig:tcom_box_and_tcom_table} b) confirm that \ac{IPC} significantly reduces communication latency. 
For example,  FastDDS best-effort median $T_{Com}$ in the Camera 1-1 scenario decreased from \SI{35.328}{\milli\second} (D1-1) to \SI{1.714}{\milli\second} (C1-1) when switching to \ac{IPC}, while Zenoh exhibited a similar reduction from \SI{38.528}{\milli\second} (D1-1) to \SI{3.774}{\milli\second} (C1-1). 
vSomeIP showed less improvement, reducing latency from \SI{56.225}{\milli\second} (D1-1) to \SI{22.979}{\milli\second} (C1-1), possibly due to its reliable communication mode. 
Large-message transmissions benefited most notably from \ac{IPC}. 
For IMU messages, latency improvements were less pronounced, with vSomeIP producing results comparable to FastDDS and Zenoh.
This behavior was consistent across all tested topologies.

\subsection{Message Loss}
We also measured the message loss across our experiments between communicating applications to determine the reliability of middlewares.
The results of these experiments are shown in \cref{fig:scaling_and_mem}.

\subsubsection{\textbf{Finding 7: Zenoh performs more robustly, while FastDDS fails to transmit in D3-1 and D2-2 topologies under high data rates}}
\label{sec:results:finding_7}
FastDDS, with its default configuration, fails to transmit \SI{4}{\mega\byte} Camera and Lidar messages in the D3-1 topology and succeeds only with best-effort transport in the D2-2 topology.  
In these cases, all messages are lost. This issues occurs repeatedly and predictably in transmission to the receivers
In \cref{fig:tcom_box_and_tcom_table}~a), these transmission failures are indicated by missing box plots.  
vSomeIP and Zenoh both reliably managed message transmissions in these cases, although with generally higher latency in the case of vSomeIP.   
This poor performance by FastDDS is likely related to the high data rates of our Lidar and Camera message type. 
In fact, investigating this issue, we were able to achieve successful transmissions in these configurations by reducing either the frequency or the message size.
This issue is likely caused by UDP or Ethernet frame segmentation at high data rates as described in its documentation\footnote{\url{https://fast-dds.docs.eprosima.com/en/3.1.x/fastdds/use_cases/large_data/large_data.html}}.

\subsubsection{\textbf{Finding 8: vSomeIP and FastDDS exhibit message loss in D3-1 and D2-2 topologies}}
\label{sec:results:finding_8}
\begin{figure*}[!htb]
    \centering
    \subfloat[]{\includegraphics[width=0.95\textwidth]{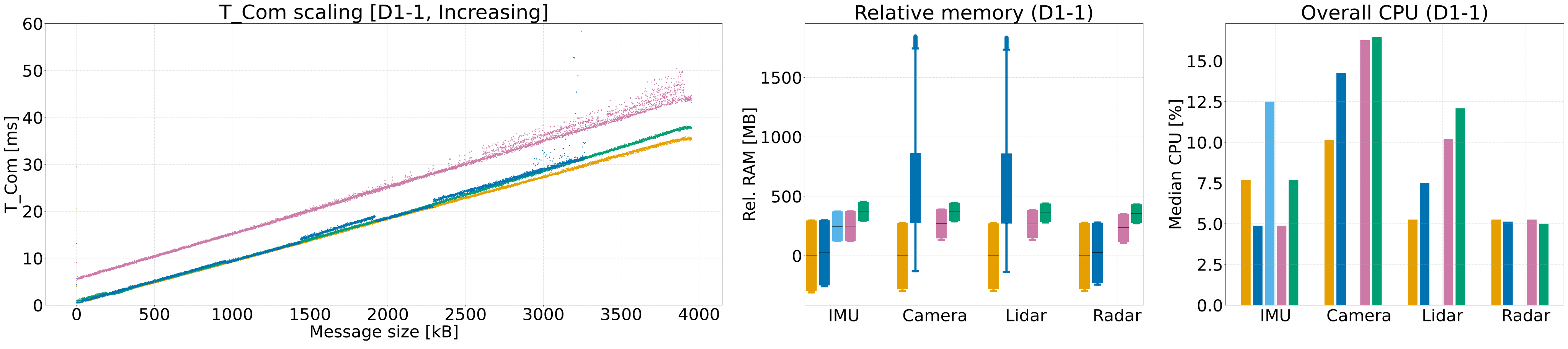}}
    \vspace{0.1ex}
    \subfloat[]{\includeinkscape[width=0.98\textwidth]{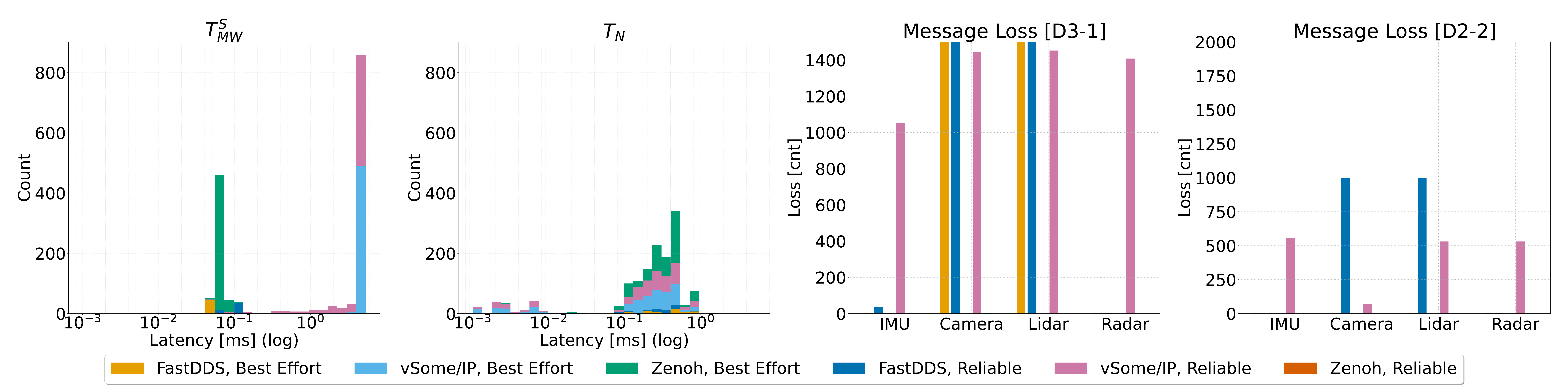}}
    \caption{(a) Left:  Scatter plot of $T_{Com}$ for steadily increasing message sizes in a D1-1 topology. (a) Right: Box plot of the RAM usage of the system relative to FastDDS best-effort as baseline and absolute median CPU usage in percent during the experiments. (b) Right: Message loss bar graph for D3-1 and D2-2 topologies, grouped by message type. The other topologies D1-1 and D1-3 experienced no message loss.(b) Left: Kernel contribution metrics $T_{MW}^{S}$ and $T_{N}$ for 1000 samples in a D1-1 topology.}
    \label{fig:scaling_and_mem}
\end{figure*}

As shown in \cref{fig:scaling_and_mem}(a), vSomeIP reliable exhibited persistent message loss in the D3-1 topology across all message sizes and experienced additional loss in D2-2 of roughly a quarter of all messages. FastDDS reliable dropped approximately half of all messages in D2-2, indicating that messages from one publisher may not have reached the shared topic. In contrast, Zenoh reliably delivered all messages in both topologies, likely because of its use of TCP transport. We omitted graphs for the D1-1 and D1-3 topologies in figure \cref{fig:scaling_and_mem} (a), as these did not show any message loss.

\subsection{Kernel Latency Contribution}
We also investigate how middleware performance is influenced by underlying kernel operations. 
To this end, we analyzed the latency in user-space and kernel-space during message transmission in a D1-1 topology with 500 samples per middleware.
The results of these experiments are shown in \cref{fig:tcom_box_and_tcom_table}.

\subsubsection{\textbf{Finding 9: Middleware systems show consistent network stack use but differ in transport defaults}}
\label{sec:results:finding_9}
All evaluated middlewares invoke the same kernel functions during transmission, mainly relying on standard UDP or TCP channels.
We observed that FastDDS uses UDP transport regardless of reliability, while vSomeIP selects UDP or TCP transport based on the configured reliability. 
Zenoh, in contrast, consistently used TCP transport even for small messages, which explains some of the differences in behavior. 
TCP transport is likely more resistant to dropped messages than UDP based transport, where the middleware has to manage retransmission. \cref{fig:scaling_and_mem} (b) seems to support this conclusion, as Zenoh performs very robustly in all configurations.

\subsubsection{\textbf{Finding 10: Kernel and network stack dominate end-to-end latency}}
\label{sec:results:finding_10}
The bulk of transmission latency in most middlewares (\textgreater50\%) originates from processes within the kernel and network stack, rather than from middleware logic.
As shown in Figure \cref{fig:tcom_box_and_tcom_table} (a) for FastDDS and Zenoh, kernel processes consistently accounted for \SIrange{0.35}{0.37} {\milli\second} (\textgreater54.8\%) of the total latency.
However, vSomeIP was an outlier, as it exhibited significantly longer kernel invocation delays, reaching up to \SI{5.10}{\milli\second} (\textgreater71.7\%) in reliable (TCP) mode.
For Zenoh and FastDDS, kernel transmission and processes form a lower bound on transmission latency, while vSOME/IPs results warrant deeper investigation.

\subsection{Scaling Behavior and Resource Usage}
Lastly, we investigate resource use and the middlewares scaling behavior with respect to message sizes and data rates.
To this end, we conducted separate scaling experiments and recorded system resource utilization during our experiments.
The results of these experiments are shown in \cref{fig:scaling_and_mem}.

\subsubsection{\textbf{Finding 11: FastDDS best-effort scales better in D1-1 communication than Zenoh and vSomeIP} } 
\label{sec:results:finding_11}
\Cref{fig:scaling_and_mem} presents $T_{Com}$ as message size increases in the D1-1 topology. The gradient of the plot visually represents our scaling factor $S$.
FastDDS best-effort demonstrates best scaling, with a scaling factor $S$ of 0.167. 
Zenoh follows closely with a $S$ of 0.171, while vSomeIP shows poorer performance with a higher factor of 0.330.
In all evaluated configurations FastDDS best-effort, Zenoh best-effort, and vSomeIP reliable successfully transmitted the full set of 5500 samples. 
Only FastDDS reliable failed to complete the full experiment, stopping at 4557 samples.
This result supports the findings in \cref{sec:results:finding_7} which showed that FastDDS reliable does not perform well for high data rates.

\subsubsection{\textbf{Finding 12: FastDDS has the lowest resource usage, while Zenoh and vSomeIP show higher median resource use}}
\label{sec:results:finding_12}
For all memory measurements, FastDDS best-effort serves as the baseline, and memory usage is reported relative to this baseline.
CPU usage is reported as absolute percentages of system usage.
FastDDS shows the lowest median memory consumption, particularly in D1-1 topologies. 
Zenoh uses \SI{374.4}{\mega\byte} more memory than FastDDS, while vSomeIP exceeds the baseline by \SI{245.6}{\mega\byte}, placing it between the two.
Median CPU usage similarly reflects this trend: Zenoh demonstrates similar or higher CPU demands than FastDDS, as shown in \cref{fig:scaling_and_mem}.
Memory usage in our best-effort configurations remains stable across message types. 
However, reliable communication in FastDDS increases the maximum memory use by up to \SI{1845.9}{\mega\byte} compared to the best-effort configuration.

\section{Conclusion}
\label{sec:conclusion}

We evaluated the performance of three middlewares for distributed real-time systems: Zenoh, FastDDS, and vSomeIP. We assessed throughput, latency, message loss, kernel latency contribution and time-to-first-message across different configurations, network topologies, and message types. 
Zenoh and FastDDS delivered comparable latency performance in most scenarios and consistently outperformed vSomeIP. 
Both Zenoh and vSomeIP handled all tested topologies and message types robustly, while FastDDS showed message loss for high data rates. 
All middleware systems benefited from centralized topology on a single device.
Zenoh and FastDDS achieved notable performance improvements, while vSomeIP benefited to a lesser extent.
Zenoh achieved the fastest time-to-first-message, closely followed by FastDDS, with vSomeIP performing significantly worse. 
In terms of resource usage, the FastDDS median resource usage was the lowest, and Zenoh was slightly less efficient.
Our investigation of kernel impact revealed that the OS network stack is the largest contributor to latency. Its performance formed a lower bound on middleware performance.

\section{Outlook}
In future work, we plan to evaluate middleware implementations for embedded platforms, the composition of the discovery process continue our investigation of kernel impact.

\section{ACKNOWLEDGMENT}
This research is accomplished within the project ”AUTOtechagil” (FKZ 01IS22088x). We acknowledge the financial support for the project by the Federal Ministry of Education and Research of Germany (BMBF).

\printbibliography

@misc{zhang_comparison_2023,
	title = {Comparison of {DDS}, {MQTT}, and {Zenoh} in {Edge}-to-{Edge} and {Edge}-to-{Cloud} {Communication} for {Distributed} {ROS} 2 {Systems}},
	author = {Zhang, Jiaqiang and Yu, Xianjia and Ha, Sier and Queralta, Jorge Pena and Westerlund, Tomi},
	month = {sep},
	year = {2023},
    archivePrefix = "arXiv",
    eprint        = "2309.07496"
}

@inproceedings{corsaro_zenoh_2023,
	title = {Zenoh: {Unifying} {Communication}, {Storage} and {Computation} from the {Cloud} to the {Microcontroller}},
	booktitle = {2023 26th {Euromicro} {Conference} on {Digital} {System} {Design} ({DSD})},
	author = {Corsaro, Angelo and Cominardi, Luca and Hecart, Olivier and Baldoni, Gabriele and Avital, Julien Enoch Pierre and Loudet, Julien and Guimares, Carlos and Ilyin, Michael and Bannov, Dmitrii},
	month = {sep},
	year = {2023},
	pages = {422--428},
    doi={10.1109/DSD60849.2023.00065}
}

@misc{eprosima_fastdds_docs,
  author       = {{eProsima}},
  title        = {{eProsima} {Fast} {DDS} {Documentation}},
  url          = {https://fast-dds.docs.eprosima.com/en/latest/},
  type         = {Documentation},
  year         = {2025},
  month        = {apr},
  urldate      = {2025-04-30}
}

@misc{autosar_ap_explanation_sw_arch,
    type = {Standard},
	title = {Explanation of {Adaptive} {Platform} {Software} {Architecture}},
	author = {{AUTOSAR Consortium}},
	month = {nov},
	year = {2024},
}

@misc{autosar_some_ip_tp,
	type = {Standard},
	title = {{Specification} on {SOME}/{IP} {Transport} {Protocol}},
	author = {{AUTOSAR Consortium}},
	month = {nov},
	year = {2024},
}

@misc{autosar_some_ip,
	type = {Standard},
	title = {{SOME}/{IP} {Protocol} {Specification}},
	author = {{AUTOSAR Consortium}},
	month = {nov},
	year = {2024},
}

@inproceedings{henle_architecture_2022,
	title = {Architecture platforms for future vehicles: a comparison of {ROS2} and {Adaptive} {AUTOSAR}},
	booktitle = {2022 {IEEE} 25th {International} {Conference} on {Intelligent} {Transportation} {Systems} ({ITSC})},
	author = {Henle, Jacqueline and Stoffel, Martin and Schindewolf, Marc and Nagele, Ann-Therese and Sax, Eric},
	month = {oct},
	year = {2022},
	pages = {3095--3102},
    doi={10.1109/ITSC55140.2022.9921894}
}

@article{wu_oops_2021,
	title = {Oops! {It}'s {Too} {Late}. {Your} {Autonomous} {Driving} {System} {Needs} a {Faster} {Middleware}},
	volume = {6},
	issn = {2377-3766},
	doi = {10.1109/LRA.2021.3097439},
	number = {4},
	journal = {IEEE Robotics and Automation Letters},
	author = {Wu, Tianze and Wu, Baofu and Wang, Sa and Liu, Liangkai and Liu, Shaoshan and Bao, Yungang and Shi, Weisong},
	month = {oct},
	year = {2021},
	pages = {7301--7308},
}

@article{kampmann_agile_2020,
	title = {Agile {Latency} {Estimation} for a {Real}-time {Service}-oriented {Software} {Architecture}},
	volume = {53},
	doi = {10.1016/j.ifacol.2020.12.1619},
	number = {2},
	journal = {IFAC-PapersOnLine},
	author = {Kampmann, Alexandru and Mokhtarian, Armin and Rogalski, Jan and Kowalewski, Stefan and Alrifaee, Bassam},
	year = {2020},
	pages = {5795--5800},
}

@inproceedings{kronauer_latency_2021,
  author={Kronauer, Tobias and Pohlmann, Joshwa and Matthé, Maximilian and Smejkal, Till and Fettweis, Gerhard},
  booktitle={2021 IEEE International Conference on Multisensor Fusion and Integration for Intelligent Systems (MFI)}, 
  title={Latency Analysis of ROS2 Multi-Node Systems}, 
  year={2021},
  volume={},
  number={},
  pages={1-7},
  doi={10.1109/MFI52462.2021.9591166}
}

@inproceedings{maruyama_exploring_2016,
	title = {Exploring the performance of {ROS2}},
	booktitle = {Proceedings of the 13th {International} {Conference} on {Embedded} {Software}},
	publisher = {ACM},
	author = {Maruyama, Yuya and Kato, Shinpei and Azumi, Takuya},
	month = {oct},
	year = {2016},
	pages = {1--10},
    doi = {10.1145/2968478.2968502}
}

@misc{kluner_modern_2024,
	title = {Modern {Middlewares} for {Automated} {Vehicles}: {A} {Tutorial}},
	shorttitle = {Modern {Middlewares} for {Automated} {Vehicles}},
	author = {Klüner, David Philipp and Molz, Marius and Kampmann, Alexandru and Kowalewski, Stefan and Alrifaee, Bassam},
	month = {dec},
	year = {2024},
    archivePrefix = "arXiv",
    eprint        = "2412.07817"
}

@article{zhu_requirements-driven_2021,
	title = {Requirements-{Driven} {Automotive} {Electrical}/{Electronic} {Architecture}: {A} {Survey} and {Prospective} {Trends}},
	volume = {9},
	journal = {IEEE Access},
	author = {Zhu, Hailong and Zhou, Wei and Li, Zhiheng and Li, Li and Huang, Tao},
	year = {2021},
    pages={100096-100112},
    doi={10.1109/ACCESS.2021.3093077}
}

@misc{mckinsey_case_nodate,
	title = {The case for an automotive software platform},
	url = {https://www.mckinsey.com/industries/automotive-and-assembly/our-insights/the-case-for-an-end-to-end-automotive-software-platform},
	urldate = {2025-04-30},
	author = {Fletcher, Ryan and Mahindroo, Abhijit and Santhanam, Nick and Tschiesner, Andreas},
}

@article{liu_impact_2022,
	title = {Impact, {Challenges} and {Prospect} of {Software}-{Defined} {Vehicles}},
	volume = {5},
	doi = {10.1007/s42154-022-00179-z},
	number = {2},
	journal = {Automotive Innovation},
	author = {Liu, Zongwei and Zhang, Wang and Zhao, Fuquan},
	month = {apr},
	year = {2022},
	pages = {180--194},
}

@misc{vsomeip,
    author = {COVESA},
	title = {{vSomeIP}: {An} {Implementation} {of} {Scalable} {Service}-{Oriented} {Middleware} {over} {IP}},
	url = {https://covesa.global/project/vsomeip/},
	urldate = {2025-04-30},
}

@inproceedings{nichitelea_automotive_2019,
	title = {Automotive {Ethernet} {Applications} {Using} {Scalable} {Service}-{Oriented} {Middleware} over {IP}: {Service} {Discovery}},
	shorttitle = {Automotive {Ethernet} {Applications} {Using} {Scalable} {Service}-{Oriented} {Middleware} over {IP}},
	doi = {10.1109/MMAR.2019.8864701},
	booktitle = {2019 24th {International} {Conference} on {Methods} and {Models} in {Automation} and {Robotics} ({MMAR})},
	author = {Nichiţelea, Teodor-Constantin and Unguritu, Maria-Geanina},
	month = aug,
	year = {2019},
	pages = {576--581},
}

@inproceedings{vetter_development_2020,
	title = {Development {Processes} in {Automotive} {Service}-oriented {Architectures}},
	booktitle = {2020 9th {Mediterranean} {Conference} on {Embedded} {Computing} ({MECO})},
	author = {Vetter, Andreas and Obergfell, Philipp and Guissouma, Houssem and Grimm, Daniel and Rumez, Marcel and Sax, Eric},
	month = jun,
	year = {2020},
	pages = {1--7},
}

@article{wang_review_2024,
	title = {Review of {Electrical} and {Electronic} {Architectures} for {Autonomous} {Vehicles}: {Topologies}, {Networking} and {Simulators}},
	volume = {7},
	shorttitle = {Review of {Electrical} and {Electronic} {Architectures} for {Autonomous} {Vehicles}},
	language = {en},
	number = {1},
	journal = {Automotive Innovation},
	author = {Wang, Wenwei and Guo, Kaidi and Cao, Wanke and Zhu, Hailong and Nan, Jinrui and Yu, Lei},
	month = feb,
	year = {2024},
	pages = {82--101},
}

@inproceedings{cai_understanding_2021,
author = {Cai, Qizhe and Chaudhary, Shubham and Vuppalapati, Midhul and Hwang, Jaehyun and Agarwal, Rachit},
title = {Understanding host network stack overheads},
year = {2021},
publisher = {Association for Computing Machinery},
booktitle = {Proceedings of the 2021 ACM SIGCOMM 2021 Conference},
pages = {65–77},
numpages = {13},
}

\end{document}